\documentclass[twocolumn,showpacs,floatfix,superscriptaddress,amsmath,amssymb,prl]{revtex4}
\usepackage{mathrsfs}
\usepackage{txfonts}
\usepackage{amssymb}
\usepackage{graphicx}
\usepackage{hyperref}
\usepackage{ulem}
 \usepackage{overpic}
 \usepackage{psfrag}
\usepackage{tabularx}
\usepackage{array}
\usepackage{placeins}
\makeatletter
\renewcommand\subsection{\@startsection
{subsection}{2}{0mm}
 {-\baselineskip}
 {0.5\baselineskip}
{\FloatBarrier\normalfont\Large\bfseries}}
\makeatother
\newcommand{\be}{\begin{equation}}
\newcommand{\ee}{\end{equation}}

\newcommand{\PreserveBackslash}[1]{\let\temp=\\#1\let\\=\temp}

\newcommand{\ket} [1] {| #1 \rangle}
\newcommand{\bra} [1] {\langle #1 |}
\newcommand{\braket}[2]{\langle #1 | #2 \rangle}

\begin{document}
\title{Finite-Size Geometric Entanglement from Tensor Network Algorithms }
\author{Qian-Qian Shi}
\affiliation{Centre for Modern Physics and Department of Physics,
Chongqing University, Chongqing 400044, The People's Republic of
China}
\author{Rom\'an Or\'us}
\affiliation{The University of Queensland, Department of Physics,
Brisbane, QLD 4072, Australia}
\author{John Ove Fj{\ae}restad}
\affiliation{The University of Queensland, Department of Physics,
Brisbane, QLD 4072, Australia}
\author{Huan-Qiang Zhou}
\affiliation{Centre for Modern Physics and Department of Physics,
Chongqing University, Chongqing 400044, The People's Republic of
China}

\begin{abstract}
The global geometric entanglement is studied in the context of
newly-developed tensor network algorithms for finite systems. For one-dimensional
quantum spin systems it is found that, at criticality,  the leading finite-size correction to the global geometric entanglement per site behaves as $b/n$,
where $n$ is the size of the system and $b$ a given coefficient. 
Our conclusion is based on the computation of the
geometric entanglement per spin for the quantum Ising model in a
transverse magnetic field and for the spin-$1/2$ XXZ model. We also discuss the possibility of coefficient $b$ being universal. 

\end{abstract}


\maketitle

 {\it Introduction.}  At zero temperature, quantum many-body systems
exhibit important collective phenomena. The complex structure of
these systems has attracted great attention in recent years. In
particular, the role played by quantum correlations (or {\it
entanglement}) in quantum phase transitions has been the subject of
numerous studies~\cite{book}. Several results for the ground states
of quantum many-body systems in one spatial dimension (1D) show that
a number of entanglement measures obey universal scaling laws close
to or at quantum critical points~\cite{first,snd,three,four,five}.
These results are, in part, due to the unique properties of the
reduced density matrices of the ground states~\cite{five}.
Complementarily, the so-called fidelity approach to quantum phase
transitions~\cite{six,sev,eight,zov} has proven specially fruitful
in determining the existence and nature of possible critical points
in quantum many-body systems.

There has also been a growing interest in investigating the distance
between the ground state of a quantum many-body system and the
closest separable state in the Hilbert space. This idea has been
quantified in terms of the so-called {\it geometric entanglement}
(GE) \cite{nine,ten}, which is a measure of the global quantum
correlations of a quantum many-body system. Remarkably, the study of
the GE in 1D many-body systems has proven fruitful because of its
connections to both the Renormalization Group and Conformal Field
Theory~\cite{four}. Moreover, the study of the GE allows to identify
separable ground states of quantum many-body systems
\cite{factorizing}. As emphasized in Refs. \cite{factorizing,libo},
the occurrence of such factorized ground states is a typical
signature of quantum phases of matter with symmetry-breaking order,
as opposed to phases with topological order where the ground state
is globally entangled. While these properties show the conceptual
significance of the GE, the GE has unfortunately also proven to be
very hard to compute both analytically and numerically. In the case of 1D systems, many
previous studies of the GE have focused on its properties for
infinite size systems, where translational invariance together with
the infinite-size limit can be fully exploited to simplify the
calculations. Its behavior for finite size 1D systems has only been analyzed in a few 
special cases \cite{ten}. 

In this paper, we show that the GE can be easily extracted for finite systems with the 
aid of newly developed Tensor Network (TN) algorithms to simulate strongly-correlated systems. 
This connection allows us to further extend the study of the GE by obtaining the finite-size
corrections to the GE for two important 1D models at
criticality: the quantum Ising and XXZ spin-$1/2$ chains. The calculations
consist of two steps: first, we compute a Matrix Product State (MPS)
representation of the ground state using TN algorithms. Second,
we determine the GE from the MPS representations thus obtained.
Our numerical observations can be summarized as follows: for a 1D system at
criticality, the finite-size corrections to the GE per site
${\mathcal E}_{n}$ for a system of size $n$ obey
 \begin{equation}\label {fit}
   {\mathcal E}_{n} \sim
   {\mathcal E}_{\infty}+\frac{b}{n} + O\left(\frac{1}{n^2}\right), ~~~n \gg 1,
 \end{equation}
with $b$ a given coefficient. As we shall see, $\mathcal{E}_n$ can 
also be interpreted as twice the free energy per site of a classical 1D vertex model. The possibility of coefficient $b$ being universal is also briefly discussed. 

{\it Global geometric entanglement per party.}
 Let us now introduce
the measure of entanglement that we shall use in this paper.
Consider a pure quantum state of $n$ parties $|\psi\rangle\in
{\mathcal H} =\bigotimes^{n}_{i=1}{\mathcal H}^{[i]}$, where
${\mathcal H}^{[i]}$ is the local Hilbert space of party $i$. The
global multipartite entanglement of $|\psi\rangle$ can be quantified
by considering the maximum fidelity $\Lambda_{{\rm max}}$ between
the quantum state $|\psi\rangle$ and all the possible separable and
normalized states $|\phi\rangle$ of the $n$ parties
 \begin{equation}\label{fid}
   \Lambda_{{\rm max} }={\rm \max_{|\phi\rangle }} \; |\langle\psi|\phi\rangle|.
 \end{equation}
  A well-defined global measure of entanglement is
obtained by taking the base-2 logarithm of $\Lambda_{{\rm
max}}^{2}$,
 \begin{equation}\label{ge}
   E(\psi)=-\log_2{\Lambda_{{\rm max}}^{2}}.
 \end{equation}
Here, we will be interested in the above quantity per party, which
is given by
\begin{equation}\label{geper}
   {\mathcal E}_{n}=n^{-1}E(\psi).
 \end{equation}
The above quantity constitutes what we call the global GE per party,
and has been investigated in different contexts
\cite{four,nine,ten,elv}. Here we study the finite-size properties of
${\mathcal E}_{n}$ as the size $n$ of the system changes. Finally,
we choose each party to be a single spin, and therefore ${\mathcal
E}_{n}$ corresponds to the GE per spin.

{\it Geometric entanglement from tensor network algorithms for 1D
systems.} For 1D quantum many-body systems, several methods have
been developed to compute their ground state properties based on MPS
representations of their wave functions. Here, we use two of these
algorithms to compute ground states of 1D quantum many-body systems
under Periodic Boundary Conditions (PBCs). These algorithms are
based on the variational optimization of the expectation value of
the energy using (i) a direct variational method~\cite{twe}, and
(ii) a variational Monte Carlo method~\cite{sandvik}. We also show
how to numerically extract the GE once the MPS representation of the
ground state wave function has been obtained. In all our derivations
we assume that the Hamiltonian for $n$ sites is made of
nearest-neighbor interactions,  $H =\sum^{n}_{i=1}h^{[i, i+1]}$,
with $n+1\equiv1$ under PBCs and $h^{[i, i+1]}$ a nearest-neighbor
two-body Hamiltonian.

Let us first recall some of the key ingredients of the direct variational algorithm from Ref. \cite{twe}
to find ground states of 1D many-body systems with PBCs.
We start from an initial guess state
$\ket{\psi_{0}}$, expressed in terms of an MPS as
\begin{equation}
\label{gues}
\ket{\psi_0}=\sum_{s_1, \cdots, s_n }{\rm Tr} \
[A^{[1]}(s_{1})A^{[2]}(s_{2})\ldots A^{[n]}(s_{n})] ~ \ket{s_1,
\cdots, s_n},
\end{equation}
where $s_i = 1, \ldots, d$ for $i = 1 \ldots , n$, and
$A^{[i]}(s_{i})$ are $D\times D$ matrices, with $d$ denoting the
dimension of the Hilbert space at every site $i$ and $D$ denoting
the so-called {\it bond dimension}. The variational optimization
algorithm finds an MPS approximation of the ground state of the
system by iteratively optimizing the expectation value of the
Hamiltonian $H$. Specifically, at every step the algorithm finds the
optimal MPS matrix at a given site. For instance, focusing on site
$m$, the expectation value of the Hamiltonian can be written as
\begin{equation}
\label{em} E=\frac{\bra{\psi} H
\ket{\psi}}{\braket{\psi}{\psi}}=\frac{\vec{A}^{[m] \dagger} ~
H_{{\rm eff}}^{m} ~ \vec{A}^{[m]}}{\vec{A}^{[m] \dagger} ~ N_{{\rm
eff}}^{m} ~ \vec{A}^{[m]}},
\end{equation}
where vector $\vec{A}^{[m]}$ has the components of matrices
$A^{[m]}(s_m)$ arranged as a vector, and $H_{{\rm eff}}^{m}$ and
$N_{{\rm eff}}^{m}$ are effective matrices that depend only on the
Hamiltonian $H$ and the MPS matrices $A^{[i]}(s_i)$ for $i \ne m$.
The update of the $d$ matrices $A^{[m]}(s_m)$ at site $m$ follows
from minimizing Eq. (\ref{em}) with respect to $\vec{A}^{[m]}$. This
amounts to solving the generalized eigenvalue problem
\begin{equation}
\label{heff} H_{{\rm eff}}^{m} ~ \vec{A}^{[m]} =\nu
~ N_{{\rm eff}}^{m} ~ \vec{A}^{[m]},
\end{equation}
with $\nu$ the expectation value of the energy. By updating the MPS
matrices site by site, the approximate ground state is obtained as
the energy $E$ monotonically decreases.

Next, we review the basics of the variational quantum Monte Carlo
optimization algorithm (VQMC) from Ref.~\cite{sandvik}. This
algorithm takes advantage of translational invariance as well as of
other symmetries in the system to accelerate convergence. More
specifically, for a translationally invariant system, the MPS of its
ground state is described by just a single site-independent set of
$d$ matrices, that is, $A^{[i]}(s_i)\equiv A(s) ~ \forall i$.
Furthermore, for systems invariant under reflection symmetry, the
condition $A^{T}\equiv A$ is imposed. The algorithm starts by giving
random values to the components of matrices $A(s)$. Then, these
components are changed in order to minimize the ground state energy.
The changes in the components $A(s)_{\alpha \beta}$ at step $k$ in
the algorithm follow the rule: $
  A(s)_{\alpha \beta} \rightarrow A(s)_{\alpha \beta} - \delta(k) \cdot r(s)_{\alpha \beta} \cdot {\rm sign} (\partial E/\partial
  A(s)_{\alpha \beta}),
$ where $r(s)_{\alpha \beta}$ is some random number between $0$ and
$1$, and $\delta(k)$ is the maximum allowed variation at step $k$
(which is set to monotonically decrease with $k$). Thus, each
component $A(s)_{\alpha \beta}$ is modified independently by a
random and well bounded variation, and the acceptance or rejection
of this variation follows a usual Monte Carlo rule. Also, as the
expectation value of the energy decreases and converges, more
sampling is required to avoid undesirable noise effects. If the
different parameters of the algorithm are chosen properly, then the
expectation value of the energy decreases stably during the updating
process. Notice that the particular value of the parameters used in
the procedure also affect the speed and precision of the
convergence. Let us also mention that the VQMC algorithm has a more favorable 
scaling with the MPS parameters than the direct variational algorithm. 

Once the MPS for the ground state is obtained using any of the above
two methods, the GE can be computed according to its definition. For
a system of $n$ spins $1/2$, we maximize Eq. (\ref{fid}) with
respect to the separable states $\ket{\phi}$ of $n$ spins,
$\ket{\phi}=\bigotimes_{i=1}^{n}(\cos(\xi_{i})\ket{0}
+\sin(\xi_{i})\ket{1})$, where $\ket{0}$ and $\ket{1}$ are the
eigenstates of the $S_z$ spin operator with eigenvalue $\pm 1/2$
respectively. For the translationally invariant (TI) Hamiltonians that we
consider in this work, we assume that the maximization of Eq.
(\ref{fid}) can be done with respect to the set of separable states
that are also TI, that is, we assume that
$\xi_{i}\equiv\xi~\forall i$~\cite{footnote}. Let us further justify this assumption: the Hamiltonians that we consider in this paper favor a ferromagnetic alignment of the spins in the quantum system, that is, spins tend to be parallel to each other. This simplifies the calculation of the geometric entanglement in that it is possible to optimize over the family of TI product states with a period of one site. Notice that if the Hamiltonans were such that the spins would tend to align antiferromagnetically, then the geometric entanglement should be obtained by optimizing over a broader family of states (most probably over those product states that are TI with a period of two sites). As we shall see, this is not the case in our work. In fact, for systems of small size, and for the Hamiltonians that we will consider here, we have numerically checked that the GE obtained by optimizing over the family of TI product states with period of one site is equal to that obtained by optimizing over the family of TI product states with period of two sites (see the Appendix), which corroborates the validity of our assumption.  

As such, the fidelity
$\Lambda(\xi)$ between the separable state $\ket{\phi}$ and the MPS
approximation to the ground state $\ket{\psi_{g}}$ takes the form
\begin{equation}
\label{lambda}
\Lambda(\xi)=\left| \braket{\psi_{g}}{\phi}\right|=
\Big|{\rm Tr}  \left( T^{[1]} T^{[2]} \cdots T^{[n]} \right) \Big|,
\end{equation}
where $T^{[i]} = \sum_{s_i} A_g^{[i]}(s_i) \otimes B(s_i)$ are
zero-dimensional transfer matrices $\forall i$ of some 1D classical
vertex model, with  $B(0)=\cos(\xi), B(1)=\sin(\xi)$, and
$A^{[i]}_{g}(s_i)$ the matrices at site $i$ in the MPS
representation of the ground state wave function $\ket{\psi_{g}}$.
Eq. (\ref{lambda}) can be further simplified if the matrices
$A^{[i]}(s_i)$ are site-independent. In this case we have that
$T^{[i]} \equiv T ~ \forall i$ and therefore the fidelity can be
expressed as $\Lambda(\xi)= \big|{\rm
Tr}(T^{n})\big|=\Big|\sum_{a}(\lambda_a)^{n}\Big|$, where
$\lambda_{a}$ are the eigenvalues of transfer matrix $T$. Finally,
the maximization of the fidelity $\Lambda(\xi)$ with respect to the
angle $\xi$ can be achieved by means of standard numerical
optimization algorithms, and from here the GE per spin
$\mathcal{E}_n$ readily follows from Eqs. (\ref{ge}) and
(\ref{geper}). Importantly, $\mathcal{E}_n$ can be understood as
twice the free energy per site of the 1D classical vertex model with
partition function $\Lambda(\xi_{max})$, where $\xi_{max}$ is the
maximizing angle.

{\it Models.} We focus on the quantum Ising spin chain with
transverse magnetic field, and on the XXZ quantum spin chain. The
Hamiltonian which accommodates both models as two limiting cases
takes the form
\begin{equation}
\small
  H=-\sum_{{i=1}}^{n}\left(\frac{1+\gamma}{2}S^{[i]}_{x}S^{[i+1]}_{x}+
  \frac{1-\gamma}{2}S^{[i]}_{y}S^{[i+1]}_{y}+\frac {\Delta}{2} S^{[i]}_{z}S^{[i+1]}_{z}+ \lambda
  S^{[i]}_{z}\right), \label{ham}
\end{equation}
where $S^{[i]}_{\alpha}$ ($\alpha=x,y,z$) are the spin operators of
the $i$-th spin $1/2$, and parameters $\Delta$, $\gamma$ and
$\lambda$ characterize different aspects of the model (anisotropies,
magnetic field). The Hamiltonian from Eq. (\ref{ham}) exhibits very
rich physics. Here, we consider the following two cases:

(i) For $\gamma=1$ and $\Delta=0$, the model corresponds to the 1D
quantum Ising model in a transverse magnetic field, with a critical
point at $\lambda=\pm1$ in the universality class of a free Majorana
fermionic field theory. Without loss of generality, here we consider
the case $\lambda=1$.

(ii) For $\gamma=0, \lambda=0$, and $\Delta$ generic, the model
is the anisotropic XXZ model, which is critical for $\Delta \in [-1, 1]$ and in the universality class of a free compactified bosonic field theory.
The system exhibits ferromagnetic and anti-ferromagnetic
orders respectively for $\Delta > 1$ and $\Delta < -1$ \cite{fero}. In this work we
focus on the critical regime  $\Delta \in [-1, 1]$.

{\it Simulation results.} As a first test, we have compared
results for the GE obtained using both the direct variational
algorithm and the VQMC algorithm. In practice, we observe that the
direct variational algorithm seems more stable but breaks
translational invariance, while the VQMC algorithm imposes
translational invariance from the beginning but needs more efforts
to get stabilized. In order to check quantitatively both methods,
we have computed the GE for the periodic quantum Ising chain at
the critical point $\lambda =1$ for sizes $n$ from $20$ to $28$.
Fixing the precision of the ground state energy in both methods to
$10^{-6}$, we see in Table \ref{TAB1}, that the relative error of
the GE as computed by the two algorithms is always smaller than $2
\times 10^{-5}$. Thus, we conclude that both methods are
equivalent in the calculation of the GE per spin up to a small
numerical error.
\begin{table}
\begin{center}
\begin{tabular}{|c|c|c|c|c|c|}
\hline size $n$ & \hfil 20\hfil & \hfil 22\hfil \hfil & \hfil
24\hfil & \hfil 26\hfil & \hfil 28\hfil \tabularnewline[1pt]
\cline{1-6}
 direct $\mathcal{E}_n \times 10^{2}$ & \hfil  7.7469 \hfil & \hfil  7.3564 \hfil \hfil & \hfil  7.0259\hfil & \hfil  6.7441\hfil & \hfil  6.4878\hfil  \tabularnewline [1pt]\cline{1-6}
 VQMC   $\mathcal{E}_n \times 10^{2}$ & \hfil 7.7464 \hfil & \hfil 7.3563 \hfil \hfil & \hfil  7.0263 \hfil & \hfil 6.7436 \hfil & \hfil  6.4884 \hfil  \tabularnewline [0pt]\cline{1-6}
 \hline
\end{tabular}
\end{center}
\setlength{\abovecaptionskip}{5pt} \caption{The geometric
entanglement per spin $\mathcal{E}_n$ is calculated by means of both
the direct optimization and the VQMC algorithms for the quantum
Ising chain in a transverse magnetic field, at critical magnetic
field $\lambda = 1$ and system sizes $n$ from 20 to 28. Fixing the
precision of the ground state energy at $10^{-6}$, the results
obtained from both algorithms for $\mathcal{E}_n$ are comparable,
with relative differences smaller than $2 \times 10^{-5}$.}
  \label{TAB1}
\end{table}

\begin{figure}
\includegraphics[width=0.45\textwidth]{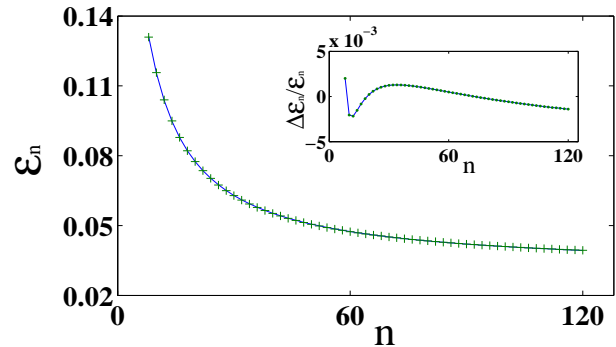}
 \caption{(color online) Main: Relation between the geometric
 entanglement per spin ${\mathcal E}_{n}$ and the chain size $n$ for the 1D quantum transverse Ising
  model with PBCs.
 For sizes $n$ from 8 to 120, the data are fitted to ${\mathcal
 E}_{n}=a+b/n+f/n^{2}$. Inset:
  The relative fitting errors $|\mathcal{E}_{n}^{{\rm fit}}-\mathcal{E}_{n}^{{\rm MPS}}|/\mathcal{E}_{n}^{{\rm MPS}}$
  are always smaller than $2.5\times10^{-3}$, where $\mathcal{E}_n^{{\rm fit}}$ is the value extracted from the fit
  and $\mathcal{E}_n^{{\rm MPS}}$ is the value extracted from the MPS algorithm for each $n$.}
  \label{FIG1}
   \end{figure}

In Fig. \ref{FIG1} we plot $\mathcal{E}_n$ for the quantum Ising
model in a transverse field at the critical point $\lambda =1$ as a
function of $n$, for sizes from $n = 8$ up to $n = 120$. Our data
suggest that there is indeed a well defined finite-size scaling law
behind the behavior of $\mathcal{E}_n$. This finite-size scaling for
the GE per spin seems to be well fitted by ${\mathcal
E}_{n}=a+b/n+f/n^{2}$, where the fitting coefficients are
$a=0.03096$, $b=1.016$, and $f=-1.713$, all with relative fitting
errors smaller than $2.5 \times 10^{-3}$. As a check, we have also
derived the thermodynamic limit of $\mathcal{E}_n$ from the exact
solution of the transverse Ising model in 1D \cite{comment}, and
obtained that at the critical point $\lambda =1$ it takes the value
${\mathcal E}_{\infty} = 0.03113$. This is comparable with the value
obtained from the extrapolation $n\rightarrow\infty$ of our fit for
$\mathcal{E}_n$, which is $a = 0.03096$. Notably, the relative error
between the exact value $\mathcal{E}_{\infty}$ and our fitted value
$a$ is of the order of $10^{-3}$.

A remark is now in order regarding potential sources of error in our fits. First, as $n$ becomes larger, it is clear that the fits should be better, therefore large system sizes are required in order to get accurate results. Second, the effects of our numerical approximation to the ground state based on matrix product states are also important. Namely, at criticality (as is the case) the introduction of a finite bond dimension in the matrix product state introduces an artificial correlation length for large system sizes (see the analysis in Ref. \cite{luca}). This, in turn, moves the ground state of the system slightly away from the true critical ground state, which implies that our results about critical properties of 1D systems can only be approximate {\it even if we were able to get the thermodynamic limit}. The fact that critical systems can only be approximately reproduced is a well-known property of all numerical algorithms based on matrix product states. Our calculations use a bond dimension of the matrix product state which is sufficiently large to assure the correctness of the geometric entanglement.

\begin{figure}
 \includegraphics[width=0.45\textwidth]{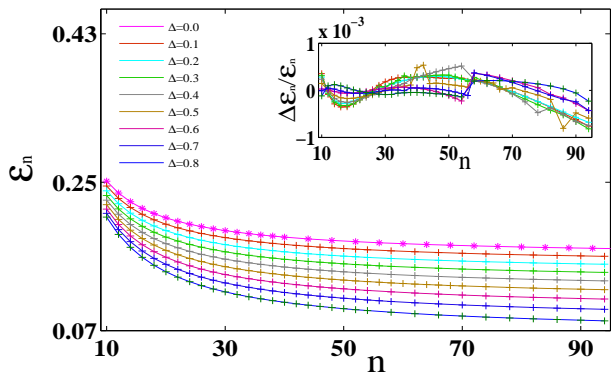}

\caption{(color online) Main:
Relation between the geometric entanglement per spin, ${\mathcal
E}_{n}$, and the chain size $n$ for the 1D XXZ model with PBCs for
sizes $n$ from 10 to 94. The plot is for nine different values of
$\Delta \ge 0$. The data are fitted to ${\mathcal
E}_{n}=a+b/n+f/n^{2}$. Inset: The relative fitting errors
$|\mathcal{E}_{n}^{{\rm fit}}-\mathcal{E}_{n}^{{\rm
MPS}}|/\mathcal{E}_{n}^{{\rm MPS}}$ are always smaller than
$10^{-3}$, where $\mathcal{E}_n^{{\rm fit}}$ is the value extracted
from the fit and $\mathcal{E}_n^{{\rm MPS}}$ is the value extracted
from the MPS algorithm for each $n$.} \label{FIG2}
\end{figure}

We have also computed $\mathcal{E}_n$ for the XXZ model with PBCs.
In the critical regime $\Delta \in [-1,1]$, we choose nine different
values $\ge 0$ of the anisotropy $\Delta$ to investigate the finite
size behavior of the GE per spin. For sizes $n$ ranging from 10 to
94 spins, ${\mathcal E}_{n}$ is again well fitted numerically by
${\mathcal E}_{n}=a+b/n+f/n^{2}$ as seen in Fig. \ref{FIG2}. In this
case, we obtain different fitting coefficients as a function of
$\Delta$, and  the relative fitting errors are always smaller than
$10^{-3}$. The specific fitting coefficients for each $\Delta$ are
given in Table \ref{TAB2}.

\begin{table}
\centering
\begin{tabular}{|r |p{2.0cm}|p{2.0cm}|p{2.0cm}|}
\hline
 $\Delta$   &\hfil $a$ \hfil & \hfil $b$\hfil & \hfil $f$\hfil \\[1pt] \cline{1-4}
  0.0 & \hfil   0.15930 \hfil&\hfil   0.9806\hfil & \hfil -0.61511 \hfil \\ [1pt] \cline{1-4}
 0.1 & \hfil   0.14922 \hfil&\hfil   1.0308\hfil & \hfil -0.68507 \hfil \\ [1pt] \cline{1-4}
 0.2 & \hfil   0.13902 \hfil&\hfil   1.0842\hfil & \hfil -0.78287 \hfil \\ [1pt] \cline{1-4}
 0.3 & \hfil   0.12858 \hfil&\hfil   1.1397\hfil & \hfil -0.86608 \hfil \\ [1pt]\cline{1-4}
 0.4 & \hfil   0.11763 \hfil&\hfil   1.2061\hfil & \hfil -0.9994 \hfil \\[1pt]\cline{1-4}
 0.5 & \hfil   0.10612 \hfil&\hfil   1.2821\hfil & \hfil -1.1607 \hfil \\[1pt] \cline{1-4}
 0.6 & \hfil   0.09383 \hfil&\hfil   1.3706\hfil & \hfil -1.3544 \hfil \\ [1pt]\cline{1-4}
 0.7 & \hfil   0.08027 \hfil&\hfil   1.4827\hfil & \hfil -1.6288 \hfil \\ [1pt]\cline{1-4}
 0.8 & \hfil   0.06469 \hfil&\hfil   1.6379\hfil & \hfil -2.0860 \hfil \\ \cline{1-4}
\hline
\end{tabular}
\setlength{\abovecaptionskip}{5pt} \caption{Fitting coefficients of
the geometric entanglement per spin, $\mathcal{E}_n$,  for nine
different values of the anisotropy $\Delta \ge 0$ in the  XXZ
model.}
  \label{TAB2}

\end{table}

An interesting issue raised by the above results is the possible universality of coefficient $b$ in Eq.(\ref{fit}). 
For the quantum Ising model, a comparison of our results with those that can be extracted from Ref.\cite{ten}
indicates that, indeed, $b$ may actually be equal to $1$ for this system.  For the XXZ model, the fact that different values of $\Delta$ correspond to different fitting coefficients could be an
indication that the finite size corrections to the GE per site at
criticality depend only on the Luttinger liquid parameter $K$ of the corresponding effective field theory at low energies, which in turn labels the corresponding universality class.  Interestingly, if $b$ were universal 
then the finite-size behavior of the GE  would be in sharp contrast to that of other quantities such as the ground state energy per site, where the universal finite-size scaling correction appears at order $O(1/n^2)$  instead of $O(1/n)$
\cite{affleck}. Notice also that other entanglement measures such as the von Neumann entropy~\cite{snd}, the single copy entanglement~\cite{three}, and the GE per region for a partition into regions of a macroscopic size \cite{four} already display universal behaviors. Despite of its inherent interest, the findings of this paper are not conclusive about the (potential) universality of coefficient $b$. This problem will be specifically addressed in a separate publication \cite{next}.  

 {\it Conclusions and discussion.}
 In this paper we have numerically investigated the finite-size GE for 1D quantum spin
 lattice systems with PBCs in the context of newly-developed TN algorithms.
 In particular, we have shown how to compute the GE using these techniques. We have also
 shown that the leading term in the finite-size correction to the GE per spin at criticality behaves as $b/n$, as evidenced by our calculations for the quantum Ising spin chain in a transverse magnetic field and for the XXZ spin chain. The possibility of this leading finite-size scaling being universal has also been discussed. 

Although we have restricted ourselves in this work to the study of
finite size quantum lattice systems in 1D, it is possible to abandon
this restriction in several ways. Specifically, one may directly
compute the GE per site for finite and infinite quantum latice
systems both in 1D and 2D by using TN algorithms \cite{iTEBD, iDMRG,
PEPS, iPEPS}. This will be the subject of future works.

{\it Acknowledgements.} The authors acknowledge very insightful comments by the referees and the editors assigned to this paper. This work is supported in part by the National Natural Science
Foundation of China (Grant Nos: 10774197 and 10874252), the
Natural Science Foundation of Chongqing (Grant No: CSTC,
2008BC2023) and the Australian Research Council.

\section{Appendix}

In this appendix we show a numerical calculation which gives evidence that the GE obtained by optimizing over the family of TI product states with period of one site is equal to that obtained by optimizing over the family of TI product states with period of two sites. More specifically, for the $XXZ$ model that we consider in this paper with $n=10$ sites a simple exact diagonalization of the Hamiltonian gives us the results from Table (\ref{TAB3}). 

\begin{table}
\centering
\begin{tabular}{|r |p{2.0cm}|p{2.0cm}|}
\hline
 $\Delta$   & \hfil GE 1-site TI  \hfil & \hfil  GE 2-site TI \hfil \\ [1pt] \cline{1-3}
   0.0   & \hfil 2.51221 \hfil & \hfil   2.51221  \hfil \\ [1pt] \cline{1-3}
   0.1   & \hfil 2.45308 \hfil & \hfil   2.45308  \hfil \\ [1pt] \cline{1-3} 
   0.2   &\hfil 2.39471 \hfil & \hfil   2.39509   \hfil \\ [1pt] \cline{1-3}
   0.3   &\hfil 2.33639 \hfil & \hfil  2.33639  \hfil \\ [1pt] \cline{1-3}
   0.4   &\hfil 2.28193  \hfil & \hfil  2.28184   \hfil \\ [1pt] \cline{1-3}
   0.5   &\hfil 2.22632  \hfil & \hfil   2.22632  \hfil \\ [1pt] \cline{1-3}
   0.6   &\hfil 2.16945  \hfil & \hfil  2.16945  \hfil \\ [1pt] \cline{1-3}
   0.7   &\hfil 2.11964 \hfil & \hfil   2.12233   \hfil \\ [1pt] \cline{1-3}
   0.8   &\hfil 2.07288 \hfil & \hfil   2.07288   \hfil \\ [1pt] \cline{1-3}
\hline
\end{tabular}
\setlength{\abovecaptionskip}{5pt} \caption{GE for the $XXZ$ model with $n=10$ sites. The results are for 1-site TI product states, and 2-site TI product states. The ground state has been obtained by an exact diagonalization of the Hamiltonian.}
\label{TAB3}
\end{table}

Remarkably, in Table (\ref{TAB3}) we can see that in both cases (1-site TI and 2-site TI) we obtain very similar results for the GE. This is an indication that, apparently, the closest product state can be chosen to be translationally invariant with period of one site, in turn being consistent with our qualitative arguments in the main body of the paper.  Notice that the above numbers also match with the results in Fig. \ref{FIG2} for $n=10$ using matrix product states. 



\begin{thebibliography}{10}
\bibitem{footnote} One may argue that there should be two parameters
in the translationally invariant factorized states instead of one
parameter $\xi$, with an extra parameter as a relative phase between
$|0 \rangle$ and $|1 \rangle$. However, such an ansatz yields the
same conclusion.
\bibitem{fero}  Notice that  the presence
of a minus sign in front of the Hamiltonian (\ref{ham}) makes our
notation different from the conventional one. As a result, the
antiferromagnetic XXZ model corresponds to $\Delta = -1$.
\bibitem{comment} Using the results in Ref.~\cite{ten}, the thermodynamic limit of the GE per spin is
${\mathcal E}_{\infty}=-\frac{2}{\ln{2}}{\rm max}
\int^{\frac{1}{2}}_{0} d \mu \ln[ \cos{\theta(\mu)}\cos^{2}{\xi/2}+
   \sin{\theta(\mu)}\sin^{2}{(\xi/2)}\cot{\pi \mu}]
$, where $\tan{2\theta(\mu)}\equiv\sin{2\pi\mu}/(1-\cos{2\pi\mu})$.
However, the results for finite size systems turn out to be
different, since the Jordan-Wigner transformation is used to solve
the model. This transformation changes the boundary conditions,
therefore affecting the GE per spin for finite size systems, but not
in the thermodynamic limit.
\end{thebibliography}
\end{document}